# Reply to "Local Filtering Fundamentally Against Wide Spectrum"

Jianwei Miao[1,2], M. C. Scott[1,2], Chien-Chun Chen[1,2], Chun Zhu[1,2], Edward R. White[1,2], Chin-Yi Chiu[2,3], B. C. Regan[1,2], Yu Huang[2,3] Laurence D. Marks[4]

[1]Department of Physics and Astronomy, [2]California NanoSystems Institute, University of California, Los Angeles, CA 90095, USA. [3]Department of Materials Science and Engineering, University of California, Los Angeles, California 90095, USA. [4]Department of Materials Science and Engineering, Northwestern University, Evanston, IL 60201, USA.

**After carefully studying the comment by Wang et al.[1], we found it includes several mistakes and unjustified statements and Wang et al. lack very basic knowledge of dislocations. Moreover, there is clear evidence indicating that Wang et al. significantly misrepresented our method and claimed something that they actually did not implement.**

First, Wang et al. simulated an image with several atoms removed (Fig. 1d), which they termed "the sharp dislocation" (quoting from their figure caption)[1]. This is not a dislocation, but consists of several vacancies (i.e. a type of point defects). In our paper[2], we claimed that, through a combination of high quality atomic resolution tilt series, centre of mass alignment, equal slope tomography (EST) reconstruction, and Fourier/Wiener filtering methods, we were able to achieve 3D imaging of the grain boundaries, stacking fault, edge and screw dislocations at atomic resolution and observe nearly all the atoms in a multiply twinned Pt nanoparticle. But we did not claim that we could image point defects inside the nanoparticle as we knew that our methods reported in the paper were not sensitive enough to detect point defects[2]. Thus, Wang et al. tried to criticize us for something that we did not claim. Furthermore, if Wang et al. want to simulate a physically realistic prismatic dislocation loop formed with a concentration of vacancies, the vacancies have to cluster on a close packed plane and collapse to form an edge dislocation loop[3]. They should then follow our procedure to perform the necessary multislice simulations if they want to verify our Fourier filtering method[2,4]. Specifically, they need to calculate a tilt series of projections using multiple slice calculations[5], conduct a 3D reconstruction with EST[2,6,7], and apply the Fourier filtering method detailed in our paper[2,4]. What they did in their comment (Figs. 1d-g) is not only an incorrect way of simulating a dislocation, but also uses a completely different simulation procedure from what we report in our paper.

Second, in Fig. 1h of their comment[1], Wang et al. show a 3D Gaussian noise background within an object support. They then applied a local filtering method with 10% cutoff of a reference Bragg peak and generated many atom-like structures (Fig. 1i). However, they failed to realize the fact that *a 3D Gaussian noise background should not produce any Bragg peaks*. Without a reference Bragg peak, our Fourier filtering method will not produce a filter. The only explanation for Wang et al.'s claim that they were able to apply our filtering method to a 3D Gaussian noise background with 10% cutoff of a reference Bragg peak and produce atom-like structures is that they have significantly misrepresented our method.

Third, their criticism of our statement – "...if all we did was simple Fourier filtering with small apertures around the Bragg spots, then this would indeed lead to artefacts; we avoided this by verifying results against unbiased Wiener filters as well as by using relatively large apertures which were adjusted to minimize signal loss"[8] – is unjustified. Our point is that for the 3D reconstruction of the experimental data, we used two different (3D Wiener and Fourier) filtering methods to independently confirm that the 3D dislocation and defect structures observed in our experimental data are real[2,4,8].

Fourth, Wang et al. stated in their comment[1] that "…as evidenced by Figure 1 which was produced after carefully following Dr. Miao's MatLab instructions shared in [3] including the aforementioned multislice simulation and other details." However, in Fig. 1, Wang et al. used the reconstruction of our experimental data, which has nothing to do with multislice simulations[5]. This statement indicates that Wang et al. claimed something that they actually did not implement.

Fifth, their statement – "In their Wiener filtering process, the noise power spectrum was estimated from the original image by averaging its Fourier spectrum over spherical shells respectively. Such averages assumed no structure in the object and spherically symmetric noise characteristics, which is generally invalid." – is incorrect. Our Wiener filtering method strictly follows the procedure published by one of us (L.D.M.) in 1996 (ref. 9). To estimate the noise level for a Wiener filter, we of course should not assume any structural information so that it is an unbiased approach.

In conclusion, there is clear evidence indicating that Wang et al. lack very basic knowledge of dislocations, significantly misrepresented our method and claimed something that they actually did not implement. Their criticisms of our paper are simply unfounded. Finally, we want to point out that we have set up two public websites where the source codes of the EST and Wiener/Fourier filtering methods as well as the data of our paper can be freely downloaded[4,7].